\documentstyle[12pt]{article}

\catcode`\@=11
\long\def\@makefntext#1{
\protect\noindent \hbox to 3.2pt {\hskip-.9pt
$^{{\ninerm\@thefnmark}}$\hfil}#1\hfill}                

 \def\@makefnmark{\hbox to 0pt{$^{\@thefnmark}$\hss}}  

\def\ps@myheadings{\let\@mkboth\@gobbletwo
\def\@oddhead{\hbox{}
\rightmark\hfil\ninerm\thepage}
\def\@oddfoot{}\def\@evenhead{\ninerm\thepage\hfil
\leftmark\hbox{}}\def\@evenfoot{}
\def\sectionmark##1{}\def\subsectionmark##1{}}

\newcounter{sectionc}\newcounter{subsectionc}\newcounter{subsubsectionc}
\renewcommand{\section}[1] {\vspace{0.6cm}\addtocounter{sectionc}{1}
\setcounter{subsectionc}{0}\setcounter{subsubsectionc}{0}\noindent
        {\bf\thesectionc. #1}\par\vspace{0.4cm}}
\renewcommand{\subsection}[1] {\vspace{0.6cm}\addtocounter{subsectionc}{1}
        \setcounter{subsubsectionc}{0}\noindent
        {\it\thesectionc.\thesubsectionc. #1}\par\vspace{0.4cm}}
\renewcommand{\subsubsection}[1]
{\vspace{0.6cm}\addtocounter{subsubsectionc}{1}
        \noindent {\rm\thesectionc.\thesubsectionc.\thesubsubsectionc.
        #1}\par\vspace{0.4cm}}

\newcounter{appendixc}
\newcounter{subappendixc}[appendixc]
\newcounter{subsubappendixc}[subappendixc]

\renewcommand{\appendix}[1] {\vspace{0.6cm}
        \refstepcounter{appendixc}
        \setcounter{figure}{0}
        \setcounter{table}{0}
        \setcounter{equation}{0}
        \renewcommand{\thefigure}{\Alph{appendixc}.\arabic{figure}}
        \renewcommand{\thetable}{\Alph{appendixc}.\arabic{table}}
        \renewcommand{\theappendixc}{\Alph{appendixc}}
        \renewcommand{\theequation}{\Alph{appendixc}.\arabic{equation}}
        \noindent{\bf Appendix \theappendixc #1}\par\vspace{0.4cm}}

\def\abstracts#1{{
        \centering{\begin{minipage}{30pc}\tenrm\baselineskip=12pt\noindent
        \centerline{\tenrm ABSTRACT}\vspace{0.3cm}
        \parindent=0pt #1
        \end{minipage}}\par}}

\renewenvironment{thebibliography}[1]
        {\begin{list}{\arabic{enumi}.}
        {\usecounter{enumi}\setlength{\parsep}{0pt}
\setlength{\leftmargin 1.25cm}{\rightmargin 0pt}
         \setlength{\itemsep}{0pt} \settowidth
        {\labelwidth}{#1.}\sloppy}}{\end{list}}

\newcounter{itemlistc}
\newcounter{romanlistc}
\newcounter{alphlistc}
\newcounter{arabiclistc}

\newcommand{\fcaption}[1]{
        \refstepcounter{figure}
        \setbox\@tempboxa = \hbox{\tenrm Fig.~\thefigure. #1}
        \ifdim \wd\@tempboxa > 6in
           {\begin{center}
        \parbox{6in}{\tenrm\baselineskip=12pt Fig.~\thefigure. #1}
            \end{center}}
        \else
             {\begin{center}
             {\tenrm Fig.~\thefigure. #1}
              \end{center}}
        \fi}

\newcommand{\tcaption}[1]{
        \refstepcounter{table}
        \setbox\@tempboxa = \hbox{\tenrm Table~\thetable. #1}
        \ifdim \wd\@tempboxa > 6in
           {\begin{center}
        \parbox{6in}{\tenrm\baselineskip=12pt Table~\thetable. #1}
            \end{center}}
        \else
             {\begin{center}
             {\tenrm Table~\thetable. #1}
              \end{center}}
        \fi}

\def\@citex[#1]#2{\if@filesw\immediate\write\@auxout
        {\string\citation{#2}}\fi
\def\@citea{}\@cite{\@for\@citeb:=#2\do
        {\@citea\def\@citea{,}\@ifundefined
        {b@\@citeb}{{\bf ?}\@warning
        {Citation `\@citeb' on page \thepage \space undefined}}
        {\csname b@\@citeb\endcsname}}}{#1}}

\newif\if@cghi
\def\cite{\@cghitrue\@ifnextchar [{\@tempswatrue
        \@citex}{\@tempswafalse\@citex[]}}
\def\citelow{\@cghifalse\@ifnextchar [{\@tempswatrue
        \@citex}{\@tempswafalse\@citex[]}}
\def\@cite#1#2{{$\null^{#1}$\if@tempswa\typeout
        {IJCGA warning: optional citation argument
        ignored: `#2'} \fi}}

\def\fnt#1#2{\footnotetext{\kern-.3em
        {$^{\mbox{\sevenrm #1}}$}{#2}}}

\font\twelvebf=cmbx10 scaled\magstep 1
 1
 1

\font\tenrm=cmr10
\font\tenit=cmti10

\font\ninerm=cmr9

\begin{document}
\newcommand{\sbar}{\,\overline{\! S}}
\newcommand{\tbar}{\overline{T}}
\newcommand{\ubar}{\overline{U}}
\newcommand{\zb}{\bar{\zeta}}
\newcommand{\psibar}{\overline{\Psi}}
\newcommand{\cm}{Commun.\ Math.\ Phys.~}
\newcommand{\pr}{Phys.\ Rev.\ D~}
\newcommand{\pl}{Phys.\ Lett.\ B~}
\newcommand{\np}{Nucl.\ Phys.\ B~}
\newcommand{\be}{\begin{equation}}
\newcommand{\en}{\end{equation}}
\begin{flushright}
\hfill{CPTH-C335.1094}\\[1mm]
NUB--3109\\[1mm]
hep-ph/9411218\\[2mm]
November 1994
\end{flushright}
\vskip 1cm
\begin{center}\twelvebf
FERMION MASSES IN SUPERSTRING THEORY\footnote{
Based on talks presented at the Spring Workshop on String Theory, ICTP,
Trieste,
11-22 April 1994, and at the Joint U.S.-Polish Workshop on Physics from Planck
Scale to Electroweak Scale, Warsaw, Poland, 21-24 September 1994.
\newline
$^\dagger$Laboratoire Propre du CNRS UPR A.0014}
\end{center}
\vspace{0.8cm}
\centerline{\tenrm I. ANTONIADIS}
\baselineskip=13pt
\centerline{\tenit
Ecole Polytechnique, Centre de Physique Th\'eorique,$^\dagger$}
\baselineskip=12pt
\centerline{\tenit 91128 Palaiseau, France}
\vspace{0.3cm}
\centerline{\tenrm and}
\vspace{0.3cm}
\centerline{\tenrm T.R. TAYLOR}
\baselineskip=13pt
\centerline{\tenit
Department of Physics, Northeastern University,}
\baselineskip=12pt
\centerline{\tenit Boston, MA 02115, U.S.A.}
\vspace{0.9cm}
\abstracts{
We give a model-independent discussion of fermion masses
in four-dimensional heterotic superstring theories. We discuss
the tree level contributions and quantum corrections, including
one-loop threshold effects and masses generated as a result of
non-perturbative supersymmetry breaking. We also point out that
superstring models give rise to a generic $\mu$-term in
the effective low energy Lagrangian.
}
\vfil
\rm\baselineskip=14pt
\setcounter{footnote}{2}
\section{OVERVIEW}
The basic assumption of superstring pheno\-me\-no\-logy
is that all light particles originate from superstring excitations
that are massless at the superstring unification scale. The hierarchy
between this high energy scale and small masses is created by the vacuum
expectations
of Higgs fields and by the supersymmetry breaking scale.
Quarks and leptons acquire masses
as a result of direct Yukawa couplings to Higgs scalars. The masses
of squarks and sleptons receive contributions
from the so-called soft terms generated by supersymmetry breaking.
What is less known, and will be discussed later in this review, is
that once local supersymmetry is broken, additional supersymmetric
mass terms can also be generated for bosons and fermions, including the
two Higgsinos of the minimal supersymmetric standard model (MSSM) and other
fermions.

According to yet another standard lore of superstring phenomenology,
supersymmetry breaking occurs as a result of non-perturbative
effects in some ``hidden sectors'' which couple to the ``observable''
world with gravitational-strength, non-renormalizable interactions only;
the soft masses are then of order of the gravitino mass,
$m_{3/2}{\propto}M_{\makebox{\tiny HID}}^3/M_{\makebox{\tiny PLANCK}}^2$.
In order for these masses to be of order 1TeV,
the scale of non-perturbative effects
must be $M_{\makebox{\tiny HID}}{\propto}10^{14}$GeV.
In typical superstring models, supersymmetry breaking is due to
gaugino condensation of a non-abelian gauge group which becomes
strong at such energy scales.

We begin by discussing fermion mass generation
due to standard Yukawa interactions that originate from explicit superpotential
terms
\begin{equation}
W= W_{ijk}Q^iQ^jh^k+s_{kl}h^kh^l+\dots, \label{spot}
\end{equation}
where $Q_{i,j}$ represent quarks or leptons, and $h^{k,l}$ the
Higgs superfields.
The ``Yukawa couplings'' $W_{ijk}$
are gauge singlet superfield functions.
They may depend on the moduli and other fields whose vacuum expectation values
(VEVs) are completely arbitrary to all orders in string perturbation theory. On
the other hand, since Higgs particles are assumed to be massless at the string
level, the functions $s_{kl}$ must have zero VEVs at the string
unification scale.

When computing the quark and lepton masses, special care must be taken
with the wave function normalization factors.
In fact, the K\"ahler potential can be expanded in powers of matter fields:
\be
K ~=~  G+ Z_{i\bar{\imath}}Q^i\bar{Q}^{\bar{\imath}}
+Z_{k\bar{k}}h^k\bar{h}^{\bar{k}}
+(H_{kl}h^kh^l+ {c.c.}) +\dots
\label{kahl}\en
The matter-independent function $G$ corresponds to the K\"ahler potential
for the moduli and other gauge singlet fields.
The wave function normalization factors for the matter fields are
determined by the functions $Z$.
The fermion mass matrix is
\begin{equation}
M_{ij}=\lambda_{ijk}\langle h^k\rangle
\label{mass}\end{equation}
where the physical Yukawa couplings
\begin{equation}
\lambda_{ijk}=e^{G/2}(Z_{i\bar{\imath}}Z_{j\bar{\jmath}}Z_{k\bar{k}})^{-1/2}
W_{ijk}
\end{equation}
and $\langle h^k\rangle$ are the VEVs of canonically normalized
Higgs fields.

The problem of computing quark and lepton masses in a given
superstring model consists of two parts.
The first part is to compute the
functions $G$, $W_{ijk}$ and $Z$ that enter into the physical Yukawa couplings.
The most important problem at this point is to determine how these
functions depend on the moduli and other gauge singlet
fields with flat potentials, since VEVs of all these fields are {\em a
priori\/}
unknown. After solving this problem, what is generally called
determination of the moduli-dependence of physical couplings,
one should compute all relevant VEVs.
Whereas the first part is of a purely kinematical nature, the second
part involves some real superstring dynamics, like gaugino condensation
and other non-perturbative phenomena.

The strategy followed in order to compute the K\"ahler potential and
the superpotential in the effective
supergravity theory of massless string excitations, is to consider
the appropriate scattering amplitudes. In this way, the tree level
quantities and one-loop corrections can be determined, as discussed
later in this review. The tree-level
superpotential (\ref{spot}) does not receive loop corrections,
in agreement with the standard non-renormalization theorems.
On the other hand, the loop expansion of the K\"ahler potential (\ref{kahl})
takes the form:
\begin{eqnarray}
G &=& -\ln(S+\sbar) + G^{(0)} + {2\over S+\sbar}G^{(1)} +\cdots \nonumber\\
Z &=& Z^{(0)} + {2\over S+\sbar}Z^{(1)} +\cdots \nonumber\\
H &=& H^{(0)} + {2\over S+\sbar}H^{(1)} +\cdots
\label{loopexp}\end{eqnarray}
Here, $S$ is the dilaton superfield which contains the dilaton as
the real part of its scalar component and the universal axion
as the imaginary part.\footnote{This axion is dual to the
two-index antisymmetric tensor.}
The dilaton VEV determines the
four-dimensional string coupling constant $g$: Re$\langle S\rangle=1/g^2$

Note that the function $H$, which mixes Higgs particles
in the K\"ahler potential (\ref{kahl}),
does not enter in the fermion mass formula
(\ref{mass}). However, once local supersymmetry is broken
by non-vanishing VEVs of some auxiliary fields, $\langle F^{\alpha}\rangle
\propto m_{3/2}$, this mixing gives rise to Higgsino masses
\begin{equation}
M_{kl}=(Z_{k\bar{k}}Z_{l\bar{l}})^{-1/2}\mu_{kl},
\label{m1}\end{equation}
where
\begin{equation}
\mu_{kl}=m_{3/2}H_{kl}-\langle\,\bar{\!
F}^{\bar{\alpha}}\rangle\partial_{\bar{\alpha}}H_{kl}.
\label{m2}\end{equation}
These masses should be interpreted as
originating from the effective low-energy superpotential term $\mu_{kl}h^kh^l$
-- the so-called $\mu$-term.

The $\mu$-term plays important role in the minimal supersymmetric
standard model.
MSSM requires the existence of two Higgs doublets, $h^1$ and $h^2$, carrying
opposite hypercharges. $h^1$ provides with masses the down quarks and leptons,
while
$h^2$ gives masses to up quarks. A superpotential term
$\mu h^1 h^2$ is necessary
in the low energy Lagrangian in order to generate masses for Higgsi\-nos
and for undesirable electro-weak axions.
In MSSM, one usually introduces by hand a parameter $\mu$ of order
of the weak scale, creating a hierarchy problem.

A class of solutions to this problem
extends the MSSM to include light singlets $s$ with Yukawa
couplings to Higgs fields, as in the second term of eq.(\ref{spot}).
This singlet could acquire a non-vanishing VEV at the electro-weak
scale, driven by the soft supersymmetry breaking, generating an effective
$\mu$-term. Similarly, a non-renormalizable effective superpotential term of
the
form
$M_{\makebox{\tiny PLANCK}}^{1-n}s^n h^1 h^2$ could be present, in which case
$\mu \propto M_{\makebox{\tiny PLANCK}}^{1-n}\langle s
\rangle^n$, with $\langle s\rangle$ now of the order of some intermediate
scale, such as the decay constant of
an invisible axion.\cite{kn}
Another solution is to introduce K\"ahler mixing of the form $Hh^1h^2+{c.c.}$,
as in the last term of eq.(\ref{kahl}). In this case, the singlets are only
gravitationally coupled and acquire in general Planck scale VEVs. A $\mu$-term
can be then induced by local supersymmetry breaking,\cite{giud}
see eq.(\ref{m2}). Superstring theory provides a natural setting
for the latter mechanism\cite{kl} since $H$ is generically a non-vanishing
function of moduli fields, as discussed later in this review.

As it is clear from eqs.(\ref{mass}) and (\ref{m2}),
the computation of fermion masses requires
determination not only of the moduli and other scalar VEVs,
but also of their auxiliary components.\footnote{Gaugino
masses can be then computed by using standard formulas.}
This brings us back to the supersymmetry breaking problem.
It is well known that gaugino condensation generates
non-perturbative potential for the dilaton and moduli.
Below, we give a simple physical explanation of the origin
of such potentials.

In the effective low-energy Lagrangian,
the gauge kinetic terms are of the form
$\sum_{\cal G}f_{\cal G}{\cal W}_{\cal G}^2$, where ${\cal W}_{\cal G}$
are the gauge field strength superfields and the functions $f_{\cal G}$
are field-dependent gauge couplings associated with group $\cal G$:
$1/g_{\cal G}^2=\langle f_{\cal G}\rangle$.
At the tree level, these functions are universal: $f_{\cal G}=S$. This
universality is violated already at the one loop level by the
threshold corrections
which depend on the moduli as well as on the matter fields.
As an example, consider gauginos of
a pure gauge hidden sector. Non-perturbative effects give rise to a
superpotential whose magnitude is determined by the gaugino condensate
$\langle\lambda\lambda\rangle{\propto}M_{\makebox{\tiny HID}}^3{\propto}
\exp(3/2bg_{\makebox{\tiny HID}}^2)M_{\makebox{\tiny PLANCK}}^3$, where
$b$ is the one-loop beta function coefficient and
$g_{\makebox{\tiny HID}}$ is the coupling constant, $1/g_{\makebox{\tiny
HID}}^2=\langle f_{\makebox{\tiny HID}}\rangle$, of the hidden gauge group. The
fact that a non-perturbative superpotential is generated for the dilaton and
moduli is due to the dependence of the gauge coupling function
$f_{\makebox{\tiny HID}}$  on these fields. This superpotential can be derived
rigorously from the  effective Lagrangian
describing non-perturbative gaugino condensation.\cite{cond}
In some simple cases, like orbifold compactifications, explicit expressions
can be obtained for the superpotential by using symmetry arguments based on
large-small compactification radius  duality.

The gauge coupling function  $f_{\makebox{\tiny HID}}$ depends
also on the matter fields. In particular, it contains terms of the form
$M_{\makebox{\tiny PLANCK}}^{-2}\partial_k\partial_lf_{\makebox{\tiny HID}}
h^kh^l$, which depend on Higgs fields.
In this way, gaugino condensation, which gives rise to
$\langle{\cal W}_{\makebox{\tiny HID}}^2\rangle{\propto}M_{\makebox{\tiny
HID}}^3$, generates an additional contribution to the $\mu$-term, of order
of $m_{3/2}$. In other words, an explicit, supersymmetric mass term
is generated for Higgsinos by non-perturbative effects in hidden
sectors.\cite{mu,cm}  This fact is illustrated later in this review on an
orbifold example.

To summarize, there exist three basic sources of fermion masses in superstring
theory: 1) tree-level superpotential Yukawa couplings,
2) mixed terms in the K\"ahler potential which give rise to masses
once local supersymmetry is spontaneously broken,
and 3) explicit mass terms generated by non-perturbative effects like
gaugino condensation. In addition, fermion masses
can be generated by some higher weight interactions which
are briefly mentioned in section 2.

This review is organized as follows.
In the next two sections, we discuss tree-level computations of the quantities
that enter into determination of fermion masses, and illustrate
them on orbifold examples. The last two sections are devoted
to quantum corrections, illustrated again on similar orbifold examples.

\section{TREE-LEVEL RESULTS}
We will restrict our discussion to Calabi-Yau compactifications of the
heterotic superstring, in which case the underlying internal superconformal
field theory has $N=(2,2)$ world-sheet supersymmetry. In this case, the gauge
group is $E_6\times E_8$ and the matter fields transform as {\bf 27} or
$\overline{{\bf 27}}$ under $E_6$ and they are in one-to-one correspondence
with the moduli: {\bf 27}'s are related to (1,1) moduli and $\overline{{\bf
27}}$'s to (1,2) moduli.
In models with $E_6$ grand unified group, all known particles
are usually assigned to {\bf 27} representations, and Yukawa couplings
originate from the superpotential terms of the form ${\bf 27}^3$.
$E_6$ can be however broken to $SO(10)$ or to another subgroup
at the string scale; non-trivial Yukawa couplings can be then generated
between particles contained in $\overline{{\bf
27}}$'s and {\bf 27}'s. In the following discussion we will assume
that Higgs fields can originate from both {\bf 27}
and $\overline{{\bf 27}}$.
The knowledge of low-energy effective Lagrangians
is then absolutely crucial in order to identify quarks, leptons and
Higgs particles in this class of models.

The K\"ahler potential has the following power
expansion in the matter fields:
\be
K ~=~ G+A^{\alpha}A^{\bar{\alpha}}Z_{\alpha\bar{\alpha}}^{(1,1)}
+B^{\nu}B^{\bar{\nu}}
Z_{\nu{\bar\nu}}^{(1,2)}+ (A^{\alpha}B^{\nu}H_{\alpha\nu}+ c.c.) +\dots\, ,
\label{Kahler}
\en
where $A$ and $B$ refer to {\bf 27}'s and $\overline{{\bf 27}}$'s,
respectively.
The function $G$ defines the moduli metric which at the tree-level is
block-diagonal in (1,1) and (1,2) moduli: $G^{(0)} = G^{(1,1)} + G^{(1,2)}$.
The moduli metrics as well as the matter metrics $Z^{(1,1)}$ and $Z^{(1,2)}$
and the K\"ahler mixing
$H$ have been studied in the literature up to one-loop level.\cite{dkl,yuka,mu}

At the tree-level the various quantities which determine the effective low
energy $N=1$ supergravity are not independent. They are related because of the
$N=2$ world-sheet supersymmetry in the right-moving (bosonic) sector of the
heterotic superstring. An interesting consequence of the
corresponding Ward-identities is the so-called special geometry, which relates
the tree-level moduli metric to the Yukawa couplings:\cite{dkl,sg}
\be
R^{(0)}_{a\bar{c}b\bar{d}} ~=~ G^{(0)}_{a\bar{c}}G^{(0)}_{b\bar{d}}
+ G^{(0)}_{a\bar{d}}G^{(0)}_{b\bar{c}} - e^{2G^{(0)}}
W_{abe}\overline{W}_{\bar{c}\bar{d}\bar{f}}G^{(0)e\bar{f}},
\label{sg}
\en
where $R^{(0)}_{a\bar{c}b\bar{d}}$ is the Riemann tensor of the moduli
K\"ahler geometry $G^{(0)}$, and the above equation holds separately for
$(1,1)$ and $(1,2)$ moduli. Eq.(\ref{sg}) can be understood as a differential
equation which determines the moduli metric in terms of the analytic
superpotential and it can be solved in several examples.\cite{ex} On the other
hand, the tree-level matter metrics are proportional to the moduli metrics:
\begin{eqnarray}
Z^{(1,1)}_{\alpha\bar{\alpha}}&=&G^{(1,1)}_{\alpha\bar{\alpha}}\exp
(G^{(1,2)}-G^{(1,1)})/3\ , \nonumber\\
Z^{(1,2)}_{\nu\bar{\nu}}&=&G^{(1,2)}_{\nu\bar{\nu}}\exp
(G^{(1,1)}-G^{(1,2)})/3\ .
\label{mm}
\end{eqnarray}

Another consequence of $N=2$ Ward-identities is that the K\"ahler mixing
function $H$ satisfies the differential equation:\cite{mu}
\footnote{This solution may not be appropriate at enhanced symmetry
points, for instance in orbifold compactifications.}
\begin{equation}
\partial_{\bar\beta}\partial_{\bar\mu}H^{(0)}_{\alpha\nu}=
G^{(0)}_{\alpha{\bar\beta}} G^{(0)}_{\nu{\bar\mu}}\ .
\label{Hsol}
\end{equation}
The above equation can be used to identify representations
containing candidates for Higgs fields with non-trivial
K\"ahler mixings that can result in a $\mu$-term.

There is a complication which arises
in the presence of Yukawa couplings of the charged fields $A,B$ with
(non-moduli) singlets.
The interactions induced by $H_{AB}$ mix with
some other interactions, which are not described by the standard two-derivative
supergravity, corresponding to higher dimensional F-terms.\cite{mu} These new
interactions lead to additional contribution to the effective $\mu$-term.
In the globally supersymmetric limit, they have the form:
\begin{equation}
\int d^2\theta ({\bar D}^2 f^1)({\bar D}^2 f^2)\ ,
\label{hd}
\end{equation}
where $f^{1,2}$ are arbitrary functions of moduli (and singlets) and ${\bar
D}^2$ is the chiral projection. These interactions also appear as basic
building blocks in the holomorphic anomaly equations of topological amplitudes
in the heterotic case.\cite{agntv} We should note however
that they vanish in the orbifold limit.

In the presence of higher weight interactions and singlets
which couple to Higgs fields in the superpotential,
the complete mass formula becomes rather complicated.
Furthermore, Yukawa couplings of Higgs fields with
singlets produce in general a direct superpotential mass since the singlets can
acquire non-vanishing expectation values at the scale of supersymmetry
breaking.
In the case of compactifications which give rise to the particle
content of the MSSM at low energies,
there are no massless singlets coupled to Higgs particles and the above
complication does not arise. In this case, the induced $\mu$-term
depends entirely on the K\"ahler function $H$ satisfying eq.(\ref{Hsol}), as
well as on eventual non-perturbative superpotential generated at the
supersymmetry breaking scale through the matter field dependence of threshold
corrections to gauge couplings.

\section{TREE-LEVEL ORBIFOLD EXAMPLES}

Symmetric orbifolds are flat compactifications on the cotient of a
six dimensional torus over a discrete subgroup of $SU(3)$ so that one
space-time supersymmetry remains unbroken. They correspond to singular points
of Calabi-Yau manifolds with enhanced gauge symmetry $U(1)^2$, or larger. An
important property of these models is space-time duality symmetry which
contains a transformation exchanging large with small compactification radii.

For instance, for two compactified dimensions, there are four independent
parameters corresponding to the three components of the metric
$G_{IJ}$ and one component of the antisymmetric tensor
$B_{IJ}=b\varepsilon_{IJ}$. They form two complex fields $T=2({\sqrt G}+ib)$
and $U=({\sqrt G}+iG_{12})/G_{11}$ corresponding to (1,1) and (1,2) moduli,
respectively. The duality symmetry in this case forms the group of
$SL(2,Z)\times SL(2,Z)/Z_2$ transformations,
\begin{equation}
T \rightarrow {aT-ib\over icT+d},\hskip6mm
U \rightarrow {a'U-ib'\over ic'U+d'},\hskip6mm
T\leftrightarrow U ,
\label{dual}
\end{equation}
where $a,b,c,d$ are integers with $ad-bc=1$ (similarly for primed parameters).
The matter fields $\varphi$ transform under this transformations as $SL(2,Z)$
modular forms of weight $n_\varphi$:
\begin{equation}
\varphi\rightarrow (icT+d)^{-n_\varphi}\varphi\, ,
\label{phidual}
\end{equation}
and similarly under $U$ transformations. Moreover, $SL(2,Z)$ duality induces a
K\"ahler transformation under which the superpotential $W$ transforms as a
form
of weight 1:
\begin{eqnarray}
K&\rightarrow& K+\ln (icT+d)+ \ln(-ic\tbar +d)\, ,\nonumber\\[1mm]
W&\rightarrow& (icT+d)^{-1}W\, .
\label{Kdual}
\end{eqnarray}

The massless states in orbifold models fall into two sectors: (a) The untwisted
sector which contains the (1,1) and (1,2) moduli $T_\alpha$ and $U_\beta$,
in correspondence with the {\bf 27}'s $A_\alpha$ and $\overline{{\bf 27}}$'s
$B_\beta$. Here, $\alpha,\beta$ label the internal complex planes:
$\alpha=1,2,3$ while $\beta$ refers only to the $Z_2$-twisted planes, otherwise
$U$ is fixed to some background value. Also, in orbifolds with non abelian
enhanced symmetries, $T$ becomes a matrix and sums should be replaced by
traces in the subsequent formulae. (b) The twisted sector which contains matter
fields $C$'s and $\overline{C}$'s in correspondence with the blowing-up moduli
which allow to deform orbifolds into regular Calabi-Yau manifolds.

The tree-level moduli metric is:\cite{dkl}
\begin{equation}
G^{(0)}=-\sum_{\alpha=1}^3\ln(T_\alpha +\tbar_\alpha)
-\sum_\beta\ln(U_\beta +\overline{U}_\beta)\, ,
\label{modmet}
\end{equation}
while the matter metrics are:
\begin{eqnarray}
\makebox{$U$ fixed:}& &\nonumber\\
Z^{(0)}_{A\overline{A}} &=& {1\over T+\tbar},
\nonumber\\
\makebox{$Z_2$-twisted planes:}& &\nonumber\\&& \hspace*{-2.6 cm}
Z^{(0)}_{A\overline{A}} = Z^{(0)}_{B\overline{B}} =
{1\over (T+\tbar)(U+\overline{U})}, \\
Z^{(0)}_{C\overline{C}}~ =& &\hspace{-8mm}
\prod_{\alpha=1}^3{1\over(T_\alpha +\tbar_\alpha)^{n^C_\alpha}},\nonumber
\label{matmet}
\end{eqnarray}
where $T,U$ are the moduli associated to the matter fields $A,B$, and
$n^C_\alpha$ are the modular weights of $C$. Moreover, the K\"ahler mixing $H$
is nonvanishing only when the matter fields belong to the untwisted sector and
are associated with a $Z_2$-twisted internal plane;\cite{mu} in addition, it
depends on the moduli of this plane only:
\begin{eqnarray}
H^{(0)}_{AB}&=&{1\over (T+\tbar)(U+\ubar)}\ ,\nonumber\\[2mm]
H_{CC}&=&H_{CA} ~=~ H_{CB} ~=~ 0\ .
\label{hab}
\end{eqnarray}

Note that the above results are consistent with the $SL(2,Z)$ large-small
compactification radius $T$-duality (\ref{dual}-\ref{Kdual}) with the untwisted
matter fields $A$ and $B$ having modular weights 1, provided the (1,2) modulus
$U$ is not inert in the presence of matter fields:
\begin{equation}
U\rightarrow U-{ic\over icT+d}\,AB\ .
\label{utran}
\end{equation}
Similarly, $T$ transforms under $SL(2,Z)$ $U$-duality.

Once supersymmetry is broken, the induced $\mu$-term,
eq.(\ref{m2}), and a possible non-perturbative superpotential
term $W_{AB}AB$ yield the Higgsino mass:
\be
m_{AB} ~=~ m_{3/2} + (T+\tbar)F_T + (U+\ubar)F_U + (T+\tbar)
(U+\ubar)e^{G/2}W_{AB}\ ,
\label{mh}
\en
where we used the tree-level expressions (\ref{modmet}-\ref{matmet}) for
the moduli and matter metrics.

The most general superpotential,
including non-perturbative contributions,
has the following expansion in powers of matter fields:
\begin{equation}
W=W_0+W_{AB}AB+\dots
\label{wind}
\end{equation}
The gravitino mass is then $m_{3/2}=e^{G/2}W_0$.
It follows from (\ref{Kdual}) that $SL(2,Z)$ invariance
of the effective action under the transformations (\ref{dual}) and
(\ref{utran}) requires that $W_{AB}$ transforms as:
\begin{equation}
W_{AB}\! \rightarrow\!(icT+d)W_{AB}+ic\, \partial_UW_0\, .
\label{wabdual}
\end{equation}
It is remarkable this transformation property automatically implies
a non-vanishing mass term, $W_{AB}\neq 0$,
if a moduli-dependent superpotential
$W_0$ is generated. It is easy to check that the physical masses
(\ref{mh}) transform with unobservable phase factors under
the duality transformations (\ref{dual}) and (\ref{utran}).

Finally, the non vanishing superpotential terms at the trilinear level are of
the form:
\begin{equation}
W \sim A_1A_2A_3 , \, B_1B_2B_3 , \, ACC , \, BCC, \, CCC .
\label{w}
\end{equation}
The corresponding physical Yukawa couplings between three untwisted
or one untwisted and two twisted {\bf 27}'s are constants,
\begin{equation}
\lambda^{(0)}_{123}=\lambda^{(0)}_{ACC}=\lambda^{(0)}_{BCC}=g{\sqrt 2}\, ,
\label{yuka}
\end{equation}
while $\lambda^{(0)}_{CCC}$ are in general non trivial functions of the moduli
$T_\alpha$.\cite{yc}

It follows that in the context of orbifold models with Higgs fields
$h^1$ and $h^2$ contained in {\bf 27} and $\overline{{\bf 27}}$ of
the untwisted sector, there is an automatic generation of a
$\mu$-term induced by the breaking of local supersymmetry through their
mixing in the K\"ahler potential. The resulting Higgsino mass is given in
equation (\ref{mh}). Furthermore, if the top quark gets mass
as a result of trilinear superpotential couplings,
its Yukawa coupling is unified with the gauge
couplings at the string scale and is given in eq.(\ref{yuka}). This implies by
the renormalization group evolution that the top is in general heavy with a
mass close to the fixed point value. An interesting possibility is when the
whole third generation receives mass from the trilinear superpotential. In this
case, a strict prediction is obtained since all three Yukawa couplings are
equal at the unification scale: $\lambda_t=\lambda_b=\lambda_\tau=g{\sqrt 2}$.
This leads to $m_t\sim 180$ GeV together with a successful prediction for the
bottom and tau masses in the region of large $\tan\beta$ ($\sim 50$).\cite{pok}

\section{ONE-LOOP RESULTS}
Computing the loop corrections to superstring scattering amplitudes, one
integrates not only over heavy particles, but over massless particles as well.
This integration gives rise to on-shell infrared divergences,
associated with the running of low energy couplings. In the analogous
field-theoretical computations, such logarithmic divergences are usually
regulated by going off-shell, to momentum $p^2\neq 0$. It is then important to
realize that in string theory, as well as in quantum field theory, the
momentum-dependence of coupling constants is a purely infrared effect, and
therefore the corresponding $\beta$-function coefficients of the
$p^2\rightarrow 0$ divergence depend on the massless particle content
only.\cite{vt,kap}

Consider for instance the one-loop case. A generic on-shell amplitude $\cal A$
corresponding to some physical coupling of the low-energy theory is written as
an integral over the complex Teichm\"uller parameter $\tau =\tau_1 +i\tau_2$ of
the world-sheet torus inside its fundamental domain $\Gamma\equiv\{|\tau_1|\leq
\frac{1}{2},~|\tau|\geq 1\}$:
\begin{equation}
{\cal A}=\int_{\Gamma}\frac{d^2\tau}{\tau_2}B(\tau,{\bar\tau}).
\label{ampl}
\end{equation}
The presence of massless particles propagating in the loop implies that the
integrand $B$ goes to a constant $b$ as $\tau_2\rightarrow\infty$, and the
integral over the Teichm\"uller parameter diverges in the infrared. When the
logarithmic divergence is regularized and compared to the field-theoretical
$\overline{DR}$ scheme, it is converted to $b\ln(M_{st}^2/p^2)$, where
$M_{st}\simeq 5\times g \times 10^{17}$ GeV is the
string unification scale. $b$ can then be identified with the corresponding
field-theoretical $\beta$-function, while the remaining finite part of the
integral yields the moduli-dependent string threshold corrections:\cite{kap}
\begin{equation}
{\cal A}=b\ln\frac{M_{st}^2}{\mu^2} +
\int_{\Gamma}\frac{d^2\tau}{\tau_2}[B(\tau,{\bar\tau})-b].
\label{amplreg}
\end{equation}

In particular, the one-loop corrections to the K\"ahler metric $K^{(1)}$ can be
obtained by the \nolinebreak com\-pu\-tation of the one-loop
three-point amplitude involving two
complex scalars and the anti\-sym\-metric tensor field:
${\cal A}(b^{\mu\nu}\varphi(p_1)
\overline{\varphi}(p_2)) \sim \varepsilon^{\mu\nu\lambda\rho} {p_1}_\lambda
{p_2}_\rho K^{(1)}_{\varphi\overline{\varphi}}$. The result is:\cite{yuka}
\begin{equation}
K^{(1)}~=~\frac{i}{16{(2\pi)}^3}
\int_{\Gamma}\frac{d^2\tau}
{\tau_2^2}\bar{\eta}^{-2} {\rm Tr}_R F(-1)^F\ ,
\label{g1loop}
\end{equation}
where $\eta$ is the Dedekind eta-function and the trace is over the Ramond
sector of the internal $N=2$ (left-moving) superconformal field theory with
$U(1)$-charge operator $F$.

Considering the K\"ahler metric corresponding to
eq.(\ref{g1loop}), one obtains a finite correction for the moduli metric
$G^{(1)}$ (at a generic point $T\ne U$), while the integral for the matter
metric $Z^{(1)}$ is infrared divergent. The infinite part can be identified
with the one-loop anomalous dimensions of a generic $N=1$ supersymmetric field
theory, in a gauge in which the superpotential remains unrenormalized. The
remaining finite part gives the string threshold corrections to wave function
factors.\cite{yuka} These corrections determine the boundary conditions for
the physical Yukawa couplings
$\lambda_{ijk}$ at the string unification scale:
\begin{equation}
\lambda_{ijk}(M_{st})=\lambda_{ijk}^{(0)}\, [1+g^2\, (Y_i+Y_j+Y_k)]^{-1/2},
\label{yukawa}
\end{equation}
where $Y_i$ is defined as the finite part of
$Z^{(1)}_{i\bar{\imath}}/Z^{(0)}_{i\bar{\imath}}$. The one-loop corrections to
the K\"ahler mixing $H_{AB}$ can be obtained similarly by expanding
eq.(\ref{g1loop}) to first order in $AB$.

Finally, we discuss the matter field dependence of threshold corrections to
gauge couplings which lead to an additional source of Higgsino mass via
gaugino condensation. They can be obtained by the computation of the 4-point
amplitude involving two gauge bosons and two matter fields $A$ and $B$. It has
been shown\cite{yuka} that, to the leading order in matter fields,
the one-loop threshold corrections $\Delta^{\makebox{\scriptsize 1-loop}}$
satisfy
\begin{equation}
\partial_{\bar\imath} \partial_{j} \Delta^{\makebox{\scriptsize 1-loop}}
 ={\tilde b}K^{(0)}_{{\bar\imath}j}+K^{(1)}_{{\bar\imath}j}\ ,
\label{thresh}
\end{equation}
where the indices ${\bar\imath}$, $j$ represent fields which are
neutral under the gauge group associated with
$\Delta^{\makebox{\scriptsize 1-loop}}$, and $-{\tilde b}$ is
the quadratic Casimir of the adjoint representation. This result can be
anticipated on purely field-theoretical grounds:\cite{anom,ft} the first term
on the r.h.s.\ of (\ref{thresh}) is related to anomalous graphs involving the
coupling of the K\"ahler current to gauginos, whereas the second term
is due to the Green-Schwarz term which contributes to
both K\"ahler potential and gauge couplings.\cite{yuka} It turns out that in
the case of a pure gauge group with no massless matter field representations
(like $E_8$), equation (\ref{thresh}) remains valid to higher orders in matter
fields, as well.\cite{mu}

\section{ONE-LOOP ORBIFOLD EXAMPLES}
In orbifold models, the one-loop corrections to the relevant quantities which
determine the effective field theory can be explicitly calculated. For
instance, the moduli dependence of the one-loop threshold corrections to the
wave function renormalization factors of untwisted matter fields are non
vanishing only for fields associated to a $Z_2$-twisted plane,
$A$ and $B$:\cite{yuka}
\begin{eqnarray}
Y_A=Y_B &=& -{\tilde{b}}_A\{\Delta(T)+\Delta(U)\} +G^{(1)}\label{yu}\\[2mm]
\Delta(T) &{\equiv}& \ln(T+\tbar)|\eta(iT)|^4\nonumber
\end{eqnarray}
Here, $\tilde{b}_A={\hat{b}_A/ ind}$, where ${\hat{b}}_A$ is equal
to the $\beta$-function coefficient of the gauge group that transforms $A$ and
$B$ non-trivially in the corresponding $N{=}2$ supersymmetric orbifold, and
$ind$ is the index of the little subgroup of the untwisted plane in the full
orbifold group.\cite{dkl2}
In eq.(\ref{yu}) we neglected additive constants which do not
depend on the moduli $T$ and $U$. It follows that the boundary relation
(\ref{yuka}) between the untwisted Yukawa couplings and the $E_6$ gauge
coupling at the unification scale does not receive any moduli-dependent
corrections at the one-loop level:
\begin{equation}
\lambda_{123}(M_{st})=g_{E_6}(M_{st}){\sqrt{2}}\, ,
\label{exyuka}
\end{equation}
up to moduli-independent constants. In eq.(\ref{exyuka}), $g_{E_6}(M_{st})$ is
the one-loop $E_6$ gauge coupling constant at the unification scale.

The one-loop correction to the function $H_{AB}$ is:\cite{mu}
\be
H^{(1)}_{AB} ~=~ {\tilde{b}}_A \Delta_T(T) \Delta_U(U) -
{1\over U+\ubar}G^{(1)}_T
-{1\over T+\tbar}G^{(1)}_U\ ,
\label{h1AB}
\en
where subscripts on the functions $\Delta$ and $G^{(1)}$ denote partial
derivatives:
$\Delta_T(T)\equiv \partial_T\Delta(T)$ {\em etc}. The $E_8$ threshold
corrections, up to first order in the untwisted  matter fields $AB$, read:
\begin{eqnarray}
{1\over g^2_{E_8}(M_{st})} &=& {1\over g^2}
-{\tilde b}_{E_8}[ \Delta(T)+\Delta(U)]+ G^{(1)}
+ (H^{(1)}_{AB}AB+ {c.c.})\nonumber\\ &&\hspace*{-1.1cm}
+~{\tilde b}_{E_8}\{
[{1\over (T+\tbar)(U+\ubar)}
- 4\partial_T\ln\eta (iT)\partial_U\ln\eta (iU)]AB+{c.c.}\}
\label{final}
\end{eqnarray}
with ${\tilde b}_{E_8}=-30$. Note that since $G^{(1)}$ is invariant under
the transformation (\ref{dual}), the above expressions are consistent with
the invariance of the one loop correction to the K\"ahler potential,
$G^{(1)}+(H^{(1)}_{AB}AB+c.c.)$, as well as of the gauge coupling
$g^2_{E_8}(M_{st})$, under the full set of $SL(2,Z)$ duality transformations
(\ref{dual}-\ref{Kdual}) and (\ref{utran}).

We conclude by presenting an explicit expression for the Higgsino mass
(\ref{mh}) in the context of gaugino condensation. A non-perturbative
superpotential that depends on the dilaton in the right way, and satisfies the
symmetry requirements (\ref{Kdual}), (\ref{wabdual}), up to first order in the
expansion of matter fields $AB$, is:\cite{mu}
\be
W ~=~ e^{S/2{\tilde b}_{E_8}}\,\eta^{-2}(iT)\eta^{-2}(iU)
\;[1-4AB\,
\partial_T\ln\eta(iT)
\partial_U\ln\eta(iU)]\;\widetilde{W}\, ,
\label{wnonp}
\en
where $\widetilde{W}$ may depend on the moduli of the two other planes. The
second term inside the bracket, which from the point of view of
non-perturbative dynamics originates from matter field-dependent threshold
corrections (\ref{final}), gives rise to a direct mass for Higgs particles.

Using auxiliary field equations and eq.(\ref{mh}), we obtain
$F_T=-m_{3/2}\Delta_T(T)+{\cal O}(g^2)$, $F_U=-m_{3/2}\Delta_U(U) +{\cal
O}(g^2)$, and the Higgsino mass
\begin{equation}
m=-m_{3/2} (T+\tbar)(U+\ubar)\Delta_T(T)\Delta_U(U)\, ,
\label{mnonp}
\end{equation}
where we neglected terms of order  ${\cal O}(g^2)$.\\[1cm]
{\twelvebf ACKNOWLEDGEMENTS}\vskip .5cm

This work was supported in part
by the National Science Foundation under
grant PHY-93-06906, in part by the EEC contracts SC1-CT92-0792 and
CHRX-CT93-0340, and in part by a CNRS-NSF collaborative grant
INT-92-16146.\\[1cm]

\end{document}